\documentclass{ifacconf}

\usepackage{graphicx}      
\usepackage{natbib}        

\usepackage{amsmath,amsfonts}
\usepackage{algorithmic}
\usepackage{algorithm}
\usepackage{array}
\usepackage[caption=false,font=normalsize,labelfont=sf,textfont=sf]{subfig}
\usepackage{textcomp}
\usepackage{stfloats}
\usepackage{url}
\usepackage{verbatim}
\usepackage{multicol}
\usepackage{mathtools}
\usepackage{amsmath}
\usepackage{tikz}
\usepackage{pgfplots}
\usepackage{url}
\usetikzlibrary{shapes,arrows}
\usepackage{booktabs}       
\usepackage{multirow}
\usepackage{multicol}
\usepackage{graphicx}
\newtheorem{corollary}{Corollary}
\newtheorem{definition}{Definition}
\newtheorem{ex}{Example}

\newcommand{\bbN}{\mathbb{N}}

\newcommand{\bbR}{\mathbb{R}}

\newcommand{\bA}{\boldsymbol{A}}
\newcommand{\bB}{\boldsymbol{B}}
\newcommand{\bC}{\boldsymbol{C}}
\newcommand{\bD}{\boldsymbol{D}}

\newcommand{\bK}{\boldsymbol{K}}

\newcommand{\bof}{\boldsymbol{f}}
\newcommand{\bg}{\boldsymbol{g}}

\newcommand{\bi}{\boldsymbol{i}}
\newcommand{\bj}{\boldsymbol{j}}

\newcommand{\br}{\boldsymbol{r}}
\newcommand{\bs}{\boldsymbol{s}}
\newcommand{\bt}{\boldsymbol{t}}

\newcommand{\calC}{\mathcal{C}}
\newcommand{\calD}{\mathcal{D}}

\newcommand{\calZ}{\mathcal{Z}}

\DeclareMathOperator{\ceil}{ceil}

\newcommand\signals[2]{\ell_{2e}^{#2}(\bbN_0^{#1})}

\definecolor{mycolor1}{rgb}{0.00000,0.44700,0.74100}%
\definecolor{mycolor2}{rgb}{0.85000,0.32500,0.09800}%
\definecolor{mycolor3}{rgb}{0.92900,0.69400,0.12500}%
\definecolor{mycolor4}{rgb}{0.49400,0.18400,0.55600}%
\definecolor{mycolor5}{rgb}{0.6980,0.7725,0.8980}%
\definecolor{mycolor6}{rgb}{0.7686,0.8745,0.7020}%
\definecolor{mycolor7}{rgb}{1.0,0.9804,0.8039}%

\newcommand{\lsb}{[}
\newcommand{\rsb}{]}

\begin{document}
\begin{frontmatter}

\title{State space representations of the Roesser type for convolutional layers\thanksref{footnoteinfo}} 

\thanks[footnoteinfo]{This work was funded by Deutsche Forschungsgemeinschaft (DFG, German Research Foundation) under Germany's Excellence Strategy - EXC 2075 - 390740016 and under grant 468094890. The authors thank the International Max Planck Research School for Intelligent Systems (IMPRS-IS) for supporting Patricia Pauli.
}

\author[First]{Patricia Pauli} 
\author[Second]{Dennis Gramlich} 
\author[First]{Frank Allgöwer}

\address[First]{Institute for Systems Theory and Automatic Control, 
   University of Stuttgart, 70569 Stuttgart, Germany (e-mail:~\{patricia.pauli,frank.allgower\}@ist.uni-stuttgart.de).}
\address[Second]{Chair of Intelligent Control Systems, 
   Aachen, 52074 Aachen, Germany (e-mail: dennis.gramlich@ic.rwth-aachen.de)}

\begin{abstract}                
From the perspective of control theory, convolutional layers (of neural networks) are 2-D (or N-D) linear time-invariant dynamical systems. The usual representation of convolutional layers by the convolution kernel corresponds to the representation of a dynamical system by its impulse response. However, many analysis tools from control theory, e.g., involving linear matrix inequalities, require a state space representation. For this reason, we explicitly provide a state space representation of the Roesser type for 2-D convolutional layers with $c_\mathrm{in}r_1 + c_\mathrm{out}r_2$ states, where $c_\mathrm{in}$/$c_\mathrm{out}$ is the number of input/output channels of the layer and $r_1$/$r_2$ characterizes the width/length of the convolution kernel. This representation is shown to be minimal for $c_\mathrm{in} = c_\mathrm{out}$. We further construct state space representations for dilated, strided, and N-D convolutions.
\end{abstract}


\end{frontmatter}

\section{Introduction and main result}

In order to provide quick access to information relevant to the widest audience, we present our main findings and a concise introduction on the first page. Generalizations of these results, details, as well as the classification of the work in the literature can be found in the following pages.

Convolutional neural networks (CNNs) are a widely used tool for processing and classifying images \citep{li2021survey} and  convolutional layers are the major building block of these CNNs.  A 2-D convolutional layer can be defined as the mapping from an input image $(u\lsb i_1,i_2\rsb)_{i_1,i_2 \in \bbN_0}$ with $u\lsb i_1,i_2\rsb \in \bbR^{c_{\mathrm{in}}}$ to an output image $(y\lsb i_1,i_2\rsb)_{i_1,i_2 \in \bbN_0}$ with $y\lsb i_1,i_2\rsb \in \bbR^{c_{\mathrm{out}}}$ given by
\begin{align}
    y\lsb i_1,i_2\rsb = b + \sum_{t_1 =0}^{r_1}\sum_{t_2 =0}^{r_2} K\lsb t_1,t_2\rsb u \lsb i_1 - t_1,i_2-t_2\rsb . \label{eq:conv_2D}
\end{align}
Here, $u \lsb i_1 - t_1,i_2-t_2\rsb$ is tacitly set to zero for $i_1-t_1 < 0$ or $i_2-t_2 < 0$ (zero padding). The vector $b \in \bbR^{c_\mathrm{out}}$ is called bias and the collection of matrices $(K\lsb t_1,t_2\rsb)_{t_1,t_2 = 0}^{r_1,r_2}$ with $K\lsb t_1,t_2\rsb \in \bbR^{c_\mathrm{out} \times c_\mathrm{in}}$ is called convolution kernel. In the context of image processing, $u\lsb i_1,i_2\rsb$/$y\lsb i_1,i_2\rsb$ should be viewed as a pixel of an input/output image and $c_{\mathrm{in}}$/$c_{\mathrm{out}}$ should be viewed as the number of channels of the input/output image.

For the analysis of mappings \eqref{eq:conv_2D} with tools from control theory, often a state space representation is useful. As has been discussed in \citep{gramlich2023convolutional}, this is especially true for the analysis of CNNs. Therefore, we provide as our first main result a state space representation of the mapping/2-D convolution \eqref{eq:conv_2D} using the Roesser model.

\begin{thm}
    \label{thm:main1}
    The mapping \eqref{eq:conv_2D} can be represented as
    \vspace{-3pt}
    \begin{align}
		\begin{bmatrix}
			x_1\lsb i_1+1,i_2 \rsb \\
			x_2 \lsb i_1,i_2+1 \rsb  \\
			y\lsb i_1,i_2 \rsb
		\end{bmatrix}
		=
		\begin{bmatrix}
			0 & A_{11} & A_{12} & B_1\\
			  0 & A_{21} & A_{22} & B_2\\
			g & C_1 & C_2 & D
		\end{bmatrix}
		\begin{bmatrix}
			1\\
			x_1\lsb i_1,i_2 \rsb \\
			x_2\lsb i_1,i_2 \rsb\\
			u\lsb i_1,i_2 \rsb
		\end{bmatrix},
		\label{eq:RoesserSys2D}
	\end{align}
    \vspace{-2pt}
    where the matrices $A_{11},\ldots,D$ are given by
    \vspace{-1pt}
    \begin{align}\label{eq:real_2D}
        &\left[
        \begin{array}{c|c}
            A_{12} & B_1 \\ \hline
            C_2 & D
        \end{array}
        \right]=
        \left[
        \begin{array}{ccc|c}
            K\lsb r_1,r_2 \rsb & \cdots & K \lsb r_1,1 \rsb & K \lsb r_1,0 \rsb \\
            \vdots & \ddots & \vdots & \vdots \\
            K\lsb 1,r_2 \rsb & \cdots & K \lsb 1,1 \rsb & K \lsb 1,0 \rsb \\ \hline
            K\lsb 0,r_2 \rsb & \cdots & K \lsb 0,1 \rsb & K \lsb 0,0 \rsb \\
        \end{array}
        \right],\\
        &\left[
        \begin{array}{c}
            A_{11} \\ \hline
            C_1
        \end{array}
        \right]
        =
        \left[
        \begin{array}{cc}
            0 & 0\\
            I_{d_\mathrm{out}} & 0\\ \hline
            0 & I_{c_\mathrm{out}}
        \end{array}
        \right],~
        \left[
        \begin{array}{c|c}
            A_{22} & B_2
        \end{array}
        \right]
        =
        \left[
        \begin{array}{cc|c}
            0 & I_{d_\mathrm{in}} & 0\\
            0 & 0 & I_{c_\mathrm{in}}
        \end{array}
        \right],\nonumber\\
        & A_{21} = 0, ~ g = b, ~ d_\mathrm{in} = c_\mathrm{in}(r_2-1), ~ d_\mathrm{out} = c_\mathrm{out}(r_1-1). \nonumber
    \end{align}
    The state signals $(x_1\lsb i_1,i_2\rsb)_{i_1,i_2 \in \bbN_0}$ with $x_1\lsb i_1,i_2\rsb \in \bbR^{n_1}$, $n_1 = c_\mathrm{out} r_1$, and $(x_2\lsb i_1,i_2\rsb)_{i_1,i_2 \in \bbN_0}$ with $x_2\lsb i_1,i_2\rsb \in \bbR^{n_2}$, $n_2 = c_\mathrm{in} r_2$ are given inductively by \eqref{eq:conv_2D} with $x_1\lsb 0,i_2\rsb = 0$ for all $i_2\in\bbN_0$, and $x_2\lsb i_1,0\rsb = 0$ for all $i_1\in\bbN_0$.
\end{thm}

\vspace{-1pt}
The realization \eqref{eq:real_2D} has essentially been proposed in \citep{varoufakis1987minimal} for the case $c_\mathrm{in} = c_\mathrm{out} = 1$, where \eqref{eq:conv_2D} falls into the class of all-zero systems. The appeal of \eqref{eq:real_2D} is that this realization can be efficiently generated and it is minimal for important special cases.
\vspace{-1pt}
\begin{thm}
    \label{thm:main2}
    Assume $c_\mathrm{in} = c_\mathrm{out}$ and $K\lsb r_1,r_2\rsb$ has full rank. Then the state dimension $n_1+n_2 = c_\mathrm{out}r_1 + c_\mathrm{in} r_2$ is minimal among all realizations of convolutions \eqref{eq:conv_2D}.
\end{thm}
\vspace{-1pt}
Proofs for Theorems \ref{thm:main1} and \ref{thm:main2} are provided in Section \ref{sec:proofs}. 

\section{Related work and background}\label{sec:background}
Despite their widespread success in various applications that range from time series and image classification to image segmentation and face recognition \citep{li2021survey}, CNNs are hardly used in safety-critical systems due to a lack of safety and robustness guarantees. To analyze the robustness or reachability of NNs, many researchers have recently used relaxations of the underlying nonlinear activation functions to render the NN amenable to an analysis using linear matrix inequalities (LMIs) and semidefinite programming (SDP) \citep{fazlyab2020safety,fazlyab2019efficient,gramlich2023convolutional,pauli2023lipschitza,newton2023sparse,hu2020reach}. Furthermore, there has been a growing interest in LMI-based design of robust NNs \citep{pauli2021training, pauli2023lipschitz, revay2020lipschitz,revay2023recurrent}. In order to include convolutional layers in LMI-based NN analysis and design, one can translate them into sparse and repetitive fully connected layers \citep{goodfellow2016deep,aquino2022robustness,pauli2022neural}.

As \citet{gramlich2023convolutional} described, standard 2-D convolutional layers are finite impulse response (FIR) filters that can be represented as 2-D systems using the Roesser model \citep{roesser1975discrete}. Such a state space representation of a convolutional layer is a non-repetitive and more compact formulation of a convolutional layer than the formulation as a fully connected layer and, therefore, LMI-based methods have better scalability using state space representations  \citep{pauli2023lipschitza}. \citet{gramlich2023convolutional} automatically generate such representations using robust control toolboxes based on symbolic variables \citep{biannic2016generalized}, which is effective yet costly and does not scale to state-of-the-art convolutional layers. Our construction is instant, which gives it an advantage over \citet{gramlich2023convolutional} and makes it convenient to use when analyzing and synthesizing CNNs based on LMIs. What is more, state space models have recently gained importance in machine learning \citep{gu2022efficiently, gu2024mamba}.

Finding minimal state space realizations for 2-D/N-D systems, i.e., realizations of the system with minimal state dimension, is by no means trivial and has been an open problem for decades. The literature encompasses constructions of state space representations for a large number of system classes \citep{eising1978realization,eising1980state,zak1986realizations,venkataraman1994state,galkowski2005minimal,doan2015notes}, e.g., minimal realizations of all-pole and all-zero systems \citep{varoufakis1987minimal}, that are mostly based on construction by inspection using corresponding circuit implementations \citep{kung1977new,antoniou2002minimal}. In particular, convolutions are FIR filters that fall into the class of all-zero systems and \citet{varoufakis1987minimal} present a minimal state space representation for single-input-single-output all-zero systems, which conincides with our construction. We further present realizations for multiple-input-multiple-output all-zero systems that arise naturally in the context of CNNs, wherein the input size corresponds to the number of input channels and the output size corresponds to the output channels.

The current paper builds on \citet{gramlich2023convolutional} who first suggested to analyze CNNs using a 2-D systems respresentation. The main contribution of this paper is the construction of a state space representation for 2-D convolutional layers, for which we show that it is minimal for $c_\mathrm{in}=c_\mathrm{out}$. We further present state space representations for a large class of convolutional layers, including 1-D, 2-D and N-D convolutions, considering stride and dilation, and multiple input and output channels. 

The remainder of the paper is organized as follows. In Section~\ref{sec:conv}, we introduce N-D convolutional layers and state the problem and in Section~\ref{sec:proofs}, we provide proofs for our main results stated on the first page of this paper. In Section~\ref{sec:statespace}, we then present state space representations for more general convolutions and give multiple examples to illustrate the concept of our construction. Finally, in Section~\ref{sec:conclusion}, we conclude the paper.

\textbf{Notation:} An image is a sequence $(u\lsb i_1,\ldots,i_d\rsb )$ with free variables $i_1,\ldots,i_d \in \bbN_0$. In this sequence, $u\lsb i_1,\ldots,i_d\rsb$ is an element of $\bbR^c$, where $c$ is called the channel dimension (e.g., $c = 3$ for RGB images). The signal dimension $d$ will usually be $d=2$ for images or $d = 3$ for medical images. The space of such signals/sequences is denoted by $\signals{d}{c} := \{ u: \bbN_0^d \to \bbR^c\}$. Images should be understood as sequences in $\signals{d}{c}$ with a finite square as support. For convenience, we  use multi-index notation for signals, i.\,e., we denote $u\lsb i_1,\ldots,i_d\rsb$ as $u\lsb\bi \rsb$ for $\bi  \in \bbN_0^d$. For these multi-indices, we use the notation $\bi  + \bj $ for $(i_1+j_1,\ldots,i_d+j_d)$ and 
$\bi  \leq \bj $ for $i_1\leq j_1,\ldots,i_d\leq j_d$. We further denote by $[\bi ,\bj ] = \{ \bt  \in \bbN_0^d \mid \bi  \leq \bt  \leq \bj  \}$ the \emph{interval} of all multi-indices between $\bi ,\bj  \in \bbN_0^d$ and by $|[\bi ,\bj ]|$ the number of elements in this set. The interval $[\bi ,\bj [ = [\bi ,\bj -1]$. 

\section{Problem statement}\label{sec:conv}
A convolutional layer $\calC$ is a mapping from the domain $\calD_{\mathrm{in}} = \signals{d}{c_\mathrm{in}}$ to the codomain $\calD_{\mathrm{out}} = \signals{d}{c_\mathrm{out}}$ which is defined by a convolution kernel $K\lsb\bt \rsb\in\bbR^{c_\mathrm{out} \times c_\mathrm{in}}$ for $0\leq\bt <\br $ and a bias $b \in \bbR^{c_\mathrm{out}}$. It is given by
\begin{align}
    y[\bi ] = b + \sum_{0 \leq \bt  \leq \br } K[\bt ] u[\bi  - \bt ], \label{eq:conv}
\end{align}
where we tacitly set $u[\bi  - \bt ]$ to zero if $\bi  - \bt $ is not in the domain of $u\lsb\cdot\rsb$, which can be understood as zero-padding. A convolutional layer retains the dimension $d$ and the multi-index $\br  \in \bbN_0^d$ defines the size of the kernel $K\lsb\cdot\rsb$.

Our multi-index description of a convolution \eqref{eq:conv} encompasses d-D convolutions. A 1-D convolution ($d=1$) operates on a 1-D signal, e.g., a time signal, a 2-D convolution ($d=2$) operates on signals with two propagation dimensions, e.g., images, and an N-D convolution ($d=N$) considers inputs with even more input dimensions, e.g., 3-D convolutions may be used to process videos or medical images.

The problem considered in this paper is to determine a state space model of a convolutional layer \eqref{eq:conv}. Theorem~\ref{thm:main1} deals with the special case $d=2$, and in Section~\ref{sec:statespace}, we provide state space representations for extensions of  \eqref{eq:conv}. 

\vspace{5pt}
\begin{definition}[Roesser model]
	\label{def:RoesserSystem}
	A d-D system $\signals{d}{c_\mathrm{in}}\to\signals{d}{c_\mathrm{out}}, (u\lsb \bi  \rsb ) \mapsto (y\lsb \bi  \rsb )$ is described by a Roesser model
	\begin{align}
		\begin{bmatrix}
			x_1\lsb \bi  + e_1 \rsb \\
                \vdots\\
			x_{d}\lsb \bi  + e_d \rsb  \\
			y\lsb \bi  \rsb
		\end{bmatrix}
		=
		\begin{bmatrix}
			0 & A_{11} & \cdots & A_{1d} & B_1\\
                \vdots & \vdots & \ddots & \vdots & \vdots\\
			0 & A_{d1} & \cdots & A_{dd} & B_d\\
			g & C_1 & \cdots & C_d & D
		\end{bmatrix}
		\begin{bmatrix}
			1\\
			x_1\lsb \bi  \rsb \\
                \vdots\\
			x_d\lsb \bi  \rsb\\
			u\lsb \bi  \rsb
		\end{bmatrix},
		\label{eq:RoesserSys}
	\end{align}
    where $e_i$ denotes the unit vector with $1$ in the $i$-th position. Here, the collection of matrices $A_{11}$, $\ldots$, $C_d$, $D$ is called state space representation of the system, $x_1\lsb \bi  \rsb \in \bbR^{n_1}$, $\ldots$, $x_d\lsb \bi  \rsb \in \bbR^{n_d}$ are the states, $u\lsb \bi  \rsb \in \bbR^{c_\mathrm{in}}$ is the input and $y\lsb \bi  \rsb \in \bbR^{c_\mathrm{out}}$ is the output of the system. 
    We consider zero initial conditions, i.e., $x_1\lsb \bi \rsb = 0$ for all $\bi\in (\{0\}\times \bbN_0 \times \dots \times \bbN_0)$, \dots, $x_d\lsb \bi\rsb = 0$ for all $\bi\in(\bbN_0 \times \dots \times \bbN_0 \times \{0\} )$.
\end{definition}

To further simplify notation, we define
    \begin{equation*}
        \left[
        \begin{array}{c|c|c}
		\bof & \bA & \bB\\ \hline
		\bg & \bC & \bD
        \end{array}\right]:=
        \left[
        \begin{array}{c|ccc|c}
            0 & A_{11} & \cdots & A_{1d} & B_1\\
            \vdots & \vdots & \ddots & \vdots & \vdots\\
		0 & A_{d1} & \cdots & A_{dd} & B_d\\\hline
		g & C_1 & \cdots & C_d & D
	\end{array}\right].
    \end{equation*}

As shown in \citet{gramlich2023convolutional}, any 2-D convolutional layer has a realization of the form \eqref{eq:RoesserSys} and the arguments presented to prove \citep[Lemma 1]{gramlich2023convolutional} extend to N-D convolutions that can be realized as N-D systems in state space. 

\section{Proofs for Theorem \ref{thm:main1} and Theorem \ref{thm:main2}}
\label{sec:proofs}

\begin{pf}[Proof of Theorem \ref{thm:main1}]
In the following, we prove Theorem~\ref{thm:main1} by comparing the transfer functions of \eqref{eq:conv_2D} and \eqref{eq:RoesserSys2D}. Let the $z$-transform of $u\lsb i_1,i_2\rsb$ be denoted by $U(z_1,z_2)=\calZ(u\lsb i_1,i_2\rsb)$. The convolutional layer \eqref{eq:conv_2D} satisfies
\begin{equation*}
    \calZ(y \lsb i_1,i_2\rsb-b)=\widetilde{Y}(z_1,z_2)=G(z_1,z_2)U(z_1,z_2),
\end{equation*}
with the discrete-time transfer function
\begin{equation}\label{eq:tf_conv}
G(z_1,z_2)=\sum_{j_1=0}^{r_1} \sum_{j_2=0}^{r_2} K \lsb j_1,j_2 \rsb z_1^{-j_1}z_2^{-j_2}.
\end{equation}
Next, we compute the transfer function of
	\begin{align*}
		\begin{bmatrix}
			x_1\lsb i_1+1, i_2 \rsb \\
			x_{2}\lsb i_1, i_2+1 \rsb  \\
			y\lsb i_1,i_2 \rsb
		\end{bmatrix}
		=
		\begin{bmatrix}
			A_{11} & A_{12} & B_1\\
			A_{21} & A_{22} & B_2\\
			C_1 & C_2 & D
		\end{bmatrix}
		\begin{bmatrix}
			x_1\lsb i_1, i_2 \rsb \\
			x_2\lsb i_1, i_2 \rsb \\
			u\lsb i_1, i_2 \rsb
		\end{bmatrix},
	\end{align*}
as $G(z_1,z_2)=$
\begin{align}\nonumber
    &\begin{bmatrix}
        C_1 & C_2
    \end{bmatrix}
    \left(\begin{bmatrix}
        z_1 I & 0\\
        0 & z_2 I
    \end{bmatrix}
    -
    \begin{bmatrix}
        A_{11} & A_{12}\\\nonumber
        A_{21} & A_{22}
    \end{bmatrix}\right)^{-1}
    \begin{bmatrix}
        B_1\\
        B_2
    \end{bmatrix}
    + D\\\nonumber
    &=\begin{bmatrix}
        C_1 & C_2
    \end{bmatrix}
    \begin{bmatrix}
        z_1 I-A_{11} & -A_{12}\\
        0 & z_2 I-A_{22}
    \end{bmatrix}^{-1}
    \begin{bmatrix}
        B_1\\
        B_2
    \end{bmatrix}
    +D\\\nonumber
    &= C_1(z_1I-A_{11})^{-1}B_1+C_2(z_2I-A_{22})^{-1}B_2\\\nonumber
    &\quad+C_1(z_1I-A_{11})^{-1}A_{12}(z_2 I-A_{22})^{-1}B_2+D\\
    &= \sum_{j_1=0}^{r_1} \sum_{j_2=0}^{r_2} K \lsb j_1,j_2 \rsb z_1^{-j_1}z_2^{-j_2}\label{eq:tf_ss}
\end{align}
using that
\begin{equation*}
    (z_1I-A_{11})^{-1}=\begin{bmatrix}
        \frac{1}{z_1} & 0 & \dots & 0\\
        \frac{1}{z_1^2} & \frac{1}{z_1} & \ddots & \vdots\\
        \vdots & \ddots & \ddots  & 0\\
        \frac{1}{z_1^{r_1}} & \dots & \frac{1}{z_1^2} & \frac{1}{z_1} \\
    \end{bmatrix}\otimes I_{c_\mathrm{out}}
\end{equation*}
and
\begin{equation*}
    (z_2I-A_{22})^{-1}=\begin{bmatrix}
        \frac{1}{z_2} & \frac{1}{z_2^2} & \dots & \frac{1}{z_2^{r_2}}\\
        0 & \frac{1}{z_2} & \ddots & \vdots\\
        \vdots & \ddots & \ddots  & \frac{1}{z_2^2}\\
        0 & \dots & 0 & \frac{1}{z_2} \\
    \end{bmatrix}\otimes I_{c_\mathrm{in}}.
\end{equation*}
The transfer functions \eqref{eq:tf_conv} and \eqref{eq:tf_ss} are equivalent. This way, we have shown that \eqref{eq:RoesserSys2D} parameterized through \eqref{eq:real_2D} is a realization of \eqref{eq:conv_2D}.
\end{pf}

\begin{pf}[Proof of Theorem \ref{thm:main2}]
    W.l.o.g. assume $b = 0$, as it does not affect the minimal state dimension. The transfer matrix of \eqref{eq:conv_2D} is given by \eqref{eq:tf_conv}, wherein we interpret $K\lsb r_1,r_2\rsb$ as the leading matrix coefficient of this transfer matrix.

    Let a Roesser realization \eqref{eq:RoesserSys2D} of \eqref{eq:conv_2D} be given.
    We can calculate the transfer matrix of this Roesser system as
    \begin{align*}
        G(z_1,z_2) = \begin{bmatrix}
            C_1 & C_2
        \end{bmatrix}
        \begin{bmatrix}
        z_1 I - A_{11} & -A_{12}\\
        -A_{21} & z_2 I - A_{22}
        \end{bmatrix}^{-1}
        \begin{bmatrix}
            B_1\\
            B_2
        \end{bmatrix}
        +
        D.
    \end{align*}
    For $|z_1|,|z_2| > \|A\|$, this transfer function can be represented using the Neumann series as
    \begin{align*}
        \begin{bmatrix}
            C_1 z_1^{-1} & C_2 z_2^{-1}
        \end{bmatrix}
        \sum_{k=0}^\infty
        \begin{bmatrix}
        z_1^{-1} A_{11} & z_2^{-1} A_{12}\\
        z_1^{-1} A_{21} & z_2^{-1}A_{22}
        \end{bmatrix}^k
        \begin{bmatrix}
            B_1\\
            B_2
        \end{bmatrix}
        +
        D.
    \end{align*}
    Obviously, the impulse response \eqref{eq:conv_2D} and the Roesser realization \eqref{eq:RoesserSys2D} must have the same transfer function. Hence, a comparison of coefficient matrices implies that
    \begin{align*}
        \begin{bmatrix}
            C_1 z_1^{-1} & C_2 z_2^{-1}
        \end{bmatrix}
        \begin{bmatrix}
        z_1^{-1} A_{11} & z_2^{-1} A_{12}\\
        z_1^{-1} A_{21} & z_2^{-1}A_{22}
        \end{bmatrix}^k
        \begin{bmatrix}
            B_1\\
            B_2
        \end{bmatrix}
    \end{align*}
    equals $K\lsb r_1,r_2\rsb z_1^{-r_1} z_2^{-r_2}$ for $k = r-1$ and that it equals zero for $k \geq r$ for $r := r_1 + r_2$. By inserting $(z_1,z_2) = (1,1)$, we obtain that
    \begin{align*}
        \underbrace{\begin{bmatrix}
            C_1 & C_2 
        \end{bmatrix}}_{=:\bC}
        \underbrace{
        \begin{bmatrix}
            A_{11} & A_{12}\\
            A_{21} & A_{22}
        \end{bmatrix}^k}_{=:\bA^k}
        \underbrace{
        \begin{bmatrix}
            B_1\\
            B_2
        \end{bmatrix}}_{=:\bB}
    \end{align*}
    equals $K\lsb r_1,r_2\rsb$ for $k = r - 1$ and zero for $k \geq r$. Hence, 
    \begin{align*}
        \begin{bmatrix}
            \bC\\
            \bC\bA\\
            \vdots\\
            \bC\bA^{r-1}
        \end{bmatrix}
        \begin{bmatrix}
            \bA^{r-1}\bB & \cdots & \bB
        \end{bmatrix}
        \!=\!
        \begin{bmatrix}
            K\lsb r_1,r_2\rsb & * & \cdots & *\\
            0 & \ddots & \ddots & \vdots \\
            \vdots & \ddots & \ddots & *\\
            0 & \cdots & 0 & K\lsb r_1,r_2\rsb
        \end{bmatrix}\!,
    \end{align*}
    where the matrix on the right has $r$ copies of $K\lsb r_1,r_2\rsb$ on its diagonal. Since we assumed that $K\lsb r_1 , r_2 \rsb$ has full rank, this implies that the matrix on the right has rank $c_\mathrm{in}r$. Hence, the codomain of $\begin{bmatrix}
         A^{r_1+r_2-1}B & \cdots & AB & B
    \end{bmatrix}$ must have at least dimension $c_\mathrm{in}r$. This implies that $A$ must at least be an $rc_\mathrm{in} \times rc_\mathrm{in}$ matrix, i.e., the state dimension of the Roesser model must at least be $rc_\mathrm{in}$. By $r = r_1 + r_2$ and $c_\mathrm{in} = c_\mathrm{out}$ we proved of our claim.
\end{pf}

\section{General convolutions and examples}\label{sec:statespace}

In this section, we go through exemplary realizations of convolutions as 2-D systems starting from the simple 1-D convolution case and including also more general convolutions, that is N-D convolutions, strided convolutions, and dilated convolutions.

\subsection{1-D convolutions}

We can write a 1-D convolution as
\begin{align}
    y[i] = b + \sum_{t = 0}^{r} K[t] u[i - t], \label{eq:conv_1D}
\end{align}
which is a special case of \eqref{eq:conv}. Realizing 1-D convolutions in state space is straightforward. A discrete-time state space representation of the FIR filter \eqref{eq:conv_1D} is given by
	\begin{align}
		\begin{bmatrix}
			x\lsb i+1 \rsb \\
			y\lsb i \rsb
		\end{bmatrix}
		=
		\begin{bmatrix}
			0 & A & B\\
			g & C & D
		\end{bmatrix}
		\begin{bmatrix}
			1\\
			x\lsb i \rsb \\
			u\lsb i \rsb
		\end{bmatrix}
		\label{eq:RoesserSys1D}
	\end{align}
with state $x\lsb i\rsb \in\bbR^n$ and state dimension $n=rc_{\mathrm{in}}$ \citep{pauli2023lipschitza}. Here, the matrices $A$, $B$, $C$, $D$, $g$ are
\begin{align*}
    &
    & A=& \begin{bmatrix}
        0 & I_{(r-1)c_\mathrm{in}}\\
        0 & 0
    \end{bmatrix},~
    & B&= \begin{bmatrix}
        0 \\
        I_{c_\mathrm{in}} 
    \end{bmatrix},\\
    g&= b,~
    & C=&
    \begin{bmatrix}
        K[r] & \cdots & K[1]
    \end{bmatrix},~
    & D&= K[0].
\end{align*}
It is easy to verify minimality of the suggested realization by controllability and observability. The pair $(A,B)$ is controllable by design. Hence, if $(A,C)$ is observable, that is, $K\lsb r \rsb$ has full column rank, the realization is minimal. 
For $c_{\mathrm{in}} \leq c_\mathrm{out}$ most kernels render $(A,C)$ observable, while for $c_{\mathrm{in}} > c_\mathrm{out}$, the pair $(A,C)$ is not observable. 

\subsection{2-D convolutions}\label{sec:2D_conv}
Theorem~\ref{thm:main1} presents a mapping $K\mapsto(\bA,\bB,\bC,\bD)$ from the convolution kernel $K$ to a specific state space representation $(\bA,\bB,\bC,\bD)$ in the case $d = 2$. As shown in Theorem~\ref{thm:main2}, this realization is minimal in the relevant special case $c_\mathrm{in}=c_\mathrm{out}$. For general channel sizes, Theorem~\ref{thm:main1} may, however, produce state space representations that are \emph{not} minimal. 

Let us next look at a few examples of the realization presented in Theorem~\ref{thm:main1} for 2-D convolutional layers \eqref{eq:conv_2D}.

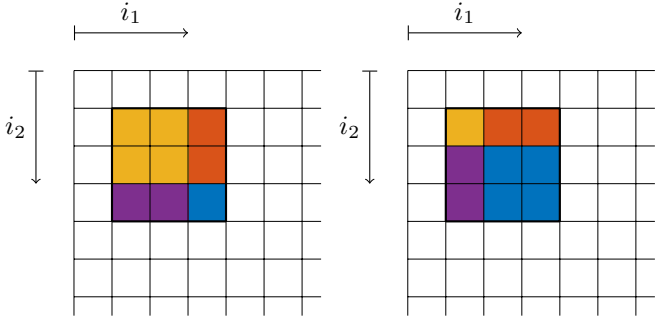
\begin{figure}
    \begin{tikzpicture}
        \draw[|->] (-0.5,2.5) -- node[above]{$i_1$} (1, 2.5);
        \draw[|->] (-1,2) -- node[left]{$i_2$} (-1, 0.5);
        \fill[mycolor1] (2/2,0/2) rectangle (3/2,1/2);
        \fill[mycolor2] (3/2,3/2) rectangle (2/2,1/2);
        \fill[mycolor3] (0/2,3/2) rectangle (2/2,1/2);
        \fill[mycolor4] (0/2,0/2) rectangle (2/2,1/2);
        \draw[step=0.5cm,black,very thin] (-1/2,-2.5/2) grid (5.5/2,4/2);
        \draw[step=1.5cm,black,thick] (0/2,0/2) grid (3/2,3/2);
    \end{tikzpicture}
    \begin{tikzpicture}
    	\draw[|->] (-0.5,2.5) -- node[above]{$i_1$} (1, 2.5);
    	\draw[|->] (-1,2) -- node[left]{$i_2$} (-1, 0.5);
        \fill[mycolor1] (1/2,0/2) rectangle (3/2,2/2);
        \fill[mycolor2] (3/2,3/2) rectangle (1/2,2/2);
        \fill[mycolor3] (0/2,3/2) rectangle (1/2,2/2);
        \fill[mycolor4] (0/2,0/2) rectangle (1/2,2/2);
        \draw[step=0.5cm,black,very thin] (-1/2,-2.5/2) grid (5.5/2,4/2);
        \draw[step=1.5cm,black,thick] (0/2,0/2) grid (3/2,3/2);
    \end{tikzpicture}
    \caption{Visualization of state space representations for $3\times 3$ kernel with stride $\bs=1$ (left) and $\bs=2$ (right).  At propagation step $i_1,i_2$, the blue pixel input enters through $D$, the purple pixel input enters through $C_2$, the red pixel input enters through $B_1$, and the yellow pixel input enters through $A_{12}$.}
    \label{fig:ss}
\end{figure}

\vspace{5pt}
\begin{ex}[Standard convolution]\label{ex:conv_3by3}
    We consider a convolutional layer \eqref{eq:conv_2D} with a $3\times 3$ kernel ($\br=2$) and stride $1$ ($\bs=1$). A state space representation is given by \eqref{eq:RoesserSys2D} with
\begin{align*}
    &\bA = \left[\begin{array}{cc|cc}
        0 & 0 & K[2,2] & K[2,1] \\
        I_{c_{\mathrm{out}}} & 0 & K[1,2] & K[1,1] \\\hline
        0 & 0 & 0      & I_{c_{\mathrm{in}}} \\
        0 & 0 & 0      & 0
    \end{array}\right],~
    &\bB& = \left[\begin{array}{c}
        K[2,0]\\
        K[1,0]\\\hline
        0\\
        I_{c_{\mathrm{in}}}
    \end{array}\right],\\
    &\bC = \left[\begin{array}{cc|cc}
        0 & I_{c_{\mathrm{out}}} &  K[0,2] & K[0,1]
    \end{array}\right],~ &\bD&=K[0,0],
\end{align*}
where the partition is by states $x_1$ and $x_2$.
\end{ex}

\vspace{5pt}
\begin{ex}[$2\times 3$ kernel]
We consider a convolutional layer \eqref{eq:conv_2D} with a $2\times 3$ kernel ($r_1=1$, $r_2=2$) and stride $1$ ($\bs=1$). A state space representation \eqref{eq:RoesserSys2D} is given by
\begin{align*}
    &\bA = \left[\begin{array}{c|cc}
        0 & K[1,2] & K[1,1]\\\hline
        0 & 0 & I_{c_\mathrm{in}}\\
        0 & 0 & 0
    \end{array}\right],~
    &\bB& = \left[\begin{array}{c}
        K[1,0]\\\hline
        0\\
        I_{c_\mathrm{in}}
    \end{array}\right],\\
    &\bC = \left[\begin{array}{c|cc}
        I_{c_\mathrm{out}} &  K[0,2] & K[0,1]
    \end{array}\right],~ &\bD&=K[0,0].
\end{align*}
\end{ex}

\vspace{5pt}
\begin{ex}[$3\times 2$ kernel]
We consider a convolutional layer \eqref{eq:conv_2D} with a $3\times 2$ kernel ($r_1=2$, $r_2=1$) and stride $1$ ($\bs=1$). A state space representation \eqref{eq:RoesserSys2D} is given by
\begin{align*}
    &\bA = \left[\begin{array}{cc|c}
        0 & 0 & K[2,1]\\
        I_{c_\mathrm{out}} & 0 & K[1,1]\\\hline
        0 & 0 & 0
    \end{array}\right],~
    &\bB& = \left[\begin{array}{c}
        K[2,0]\\
        K[1,0]\\\hline
        I_{c_\mathrm{in}}
    \end{array}\right],\\
    &\bC = \left[\begin{array}{cc|c}
        0 & I_{c_\mathrm{out}} & K[0,1]
    \end{array}\right],~ &\bD&=K[0,0].
\end{align*}
\end{ex}

Using the partition in \eqref{eq:real_2D}, the suggested state space representation is convenient to implement. We note that state space realizations are not unique, yet we suggest a specific one that is especially convenient to work with as its entries correspond to the entries of the convolution kernel. In the following, we briefly describe details of its construction. The state $x_2$ stores the pixel-wise information along the $i_2$ dimension (similar to the 1-D convolution), cmp. Fig.~\ref{fig:ss}, noting that the $x_2$ dynamic is decoupled from the $x_1$ dynamic as $A_{21}=0$. The $i_1$ dimension is then used to compactly store the pixel sum weighted by the parameter entries in $K$ for the past propagation steps in the $i_1$ direction.

    \tikzstyle{block} = [draw, fill=white, rectangle, 
    minimum height=3em, minimum width=6em]
\tikzstyle{sum} = [draw, fill=white, circle, node distance=1cm]
\tikzstyle{circ} = [draw, fill, circle, minimum size = 0.1cm, inner sep=0pt, node distance=1cm]
\tikzstyle{input} = [coordinate]
\tikzstyle{output} = [coordinate]
\tikzstyle{pinstyle} = [pin edge={to-,thin,black}]

\begin{rem}
    Representing a convolution in state space requires the choice of a propagation direction for both dimensions. Usually, for image inputs we pick the upper left corner as the origin with $i_1=i_2=0$, cmp. Fig.~\ref{fig:ss}. However, any other corner and corresponding propagation directions can also be chosen to represent the convolution equivalently. For general state space model layers the propagation dimension, i.e., time is predefined, and cannot be changed.
\end{rem}

\subsection{N-D convolutions}
Finding a mapping from $K\mapsto(\bA,\bB,\bC,\bD)$ for N-D convolutions is also possible using similar construction steps as described in Section \ref{sec:2D_conv} for 2-D convolutions. For convenience, we define the operation $\mathrm{mat}(K)$ that transforms the tensor $K$ of dimension $c_\mathrm{out}\times c_\mathrm{in} \times (r_1+1)\times \dots \times (r_d+1)$ into a 2-D matrix of dimension $c_\mathrm{out}(r_1+1)\times c_\mathrm{in} \prod_{i=2}^d (r_i+1)$. The operation $\mathrm{mat}(K)$ stacks $K[\bi ]$ in reversed order along dimension $1$ in the vertical direction of the resulting matrix and along dimensions $2$ to $d$ in the horizontal dimension of the resulting matrix, starting to stack along the $d$-th dimension, finally going to the second dimension, also in reverse order. For instance, for $d=2$ the output is
\begin{equation*}
\mathrm{mat}(K)=
\begin{bmatrix}
            K\lsb r_1,r_2 \rsb & \cdots &  K\lsb r_1,0 \rsb \\
            \vdots & \ddots & \vdots \\
            K\lsb 0,r_2 \rsb & \cdots & K \lsb 0,0 \rsb \\
\end{bmatrix}
\end{equation*}
and for $d=3$, the output is $\mathrm{mat}(K)=$
\begin{equation*}
{\small
\begin{bmatrix}
    K[r_1,r_2,r_3] & \cdots & K[r_1,r_2,0] & K[r_1,r_2-1,r_3] & \cdots & K[r_1,0,0]\\
    \vdots & \ddots & \vdots & \vdots & \ddots & \vdots \\
    K[0,r_2,r_3] & \cdots & K[0,r_2,0] & K[0,r_2-1,r_3] & \cdots & K[0,0,0]
\end{bmatrix}}.
\end{equation*}

\begin{corollary}\label{cor:min_real_ND}
The mapping \eqref{eq:conv} can be represented as \eqref{eq:RoesserSys}, where the matrices $A_{11},\ldots,D$ are given by
\begin{equation*}
\begin{split}
&\left[\begin{array}{cccc}
    A_{12} & \cdots & A_{1d} & B_1\\
    C_{2} & \cdots & C_{d} & D
\end{array}\right]= \mathrm{mat}(K),
\left[\begin{array}{ccc}
    A_{11} \\
    C_1
\end{array}\right] = 
\left[\begin{array}{ccc}
    0\\
    I_{n_1}
\end{array}\right],\\
&\left[\begin{array}{cccc}
    A_{22} & \dots & A_{2d} & B_2\\
\end{array}\right] = 
\left[\begin{array}{ccc}
    0 & I_{n_2}
\end{array}\right], \cdots,
\left[\begin{array}{ccc}
    A_{dd} & B_d\\
\end{array}\right] = 
\left[\begin{array}{ccc}
    0 & I_{n_d}
\end{array}\right],\\
& A_{21}=0,\dots, A_{d1}=0,~ A_{32}=0,\dots, A_{d2}=0,\dots,\\
& A_{dd-1}=0,~g=b,
\end{split}
\end{equation*}
    where $K[\bi ]\in\bbR^{c_\mathrm{out}\times c_\mathrm{in}},~\bi \in\lsb 0,\br \rsb$ and the kernel size is $(r_1+1)\times (r_2+1)\times\cdots\times (r_d+1)$. 
    The state, input, and output dimensions are $n_{1}=c_{\mathrm{out}}r_1, n_{2}=c_{\mathrm{in}}r_d\cdots r_2,\dots,n_{d-1}=c_{\mathrm{in}}r_dr_{d-1}, n_{d}=c_{\mathrm{in}}r_d$, $n_u=c_\mathrm{in}$, $n_y=c_\mathrm{out}$.
\end{corollary}

The proof of Corollary \ref{cor:min_real_ND} follows along the lines of the proof of Theorem~\ref{thm:main1} and is omitted. Next, we provide an example.

\vspace{5pt}
\begin{ex}[3-D convolution]\label{ex:3D_conv}
    We consider a convolutional layer \eqref{eq:conv} with $d=3$ with a $2\times 2 \times 2$ kernel and stride $1$. A state space representation is given by \eqref{eq:RoesserSys} with
\begin{align*}
    &\bA \!=\! \left[\begin{array}{c|cc|c}
        0 & K\lsb 1,1,1\rsb & K\lsb 1,1,0 \rsb & K\lsb 1,0,1 \rsb \\\hline
        0 & 0 & 0 & I_{c_\mathrm{in}}\\
        0 & 0 & 0 & 0\\\hline
        0 & 0 & 0 & 0
    \end{array}\right]\!,\\
    &\bB \!=\! \left[\begin{array}{c}
        K\lsb 1,0,0 \rsb\\\hline
        0\\
        I_{c_\mathrm{in}}\\\hline
        I_{c_\mathrm{in}}
    \end{array}\right]\!,~
    \bD = K\lsb 0,0,0\rsb,\\
    &\bC = \left[\begin{array}{c|cc|c}
        I_{c_\mathrm{out}} & K\lsb 0,1,1 \rsb & K\lsb 0,1,0 \rsb & K\lsb 0,0,1 \rsb
    \end{array}\right],
\end{align*}
where the partition is by states $x_1$, $x_2$, and $x_3$.
\end{ex}

What differs in the construction of N-D convolutions is that we store information of an (N-1)-dimensional hyperplane pixel-wise in memory and the remaining dimension stores the weighted sums of all pixels. Corresponding to the 2-D case the $x_d$ dynamic is decoupled from all other dynamics, the $x_{d-1}$ dynamic is decoupled from all other dynamics except the $x_d$ dynamic and so on, yielding a block-triangular matrix $\bA$.

\subsection{Dilated convolutions}
Dilated convolutions insert spaces between the kernel elements. The number of skipped pixels/spaces is governed by an additional hyperparameter for dilation \citep{dumoulin2016guide}. Expanding the filter allows to construct a state space representation for dilated convolutions. To illustrate the concept, we provide an example.

\vspace{5pt}
\begin{ex}[Dilated convolution] \label{ex:dilated_con}
    In case of a dilated convolution with $1$ space between the kernel elements of a $3\times3$ kernel, we expand the filter and get
    \begin{equation*}
        \bK = \begin{bmatrix}
            K[0,0] & 0 & K[0,1] & 0 & K[0,2]\\
            0 & 0 & 0 & 0 & 0\\
            K[1,0] & 0 & K[1,1] & 0 & K[1,2]\\
            0 & 0 & 0 & 0 & 0\\
            K[2,0] & 0 & K[2,1] & 0 & K[2,2]\\
        \end{bmatrix}, 
    \end{equation*}
    which we then treat as a $5\times 5$ kernel when formulating $\bA$, $\bB$, $\bC$, and $\bD$ based on \eqref{eq:real_2D}.
\end{ex}

Dilated convolutions sometimes appear in deconvolutional layers. Deconvolutional layers or transposed convolutions reconstruct the spatial dimensions of the input of the corresponding convolutional layer \citep{dumoulin2016guide}. This is achieved by dilating the convolution and zero-padding.

\subsection{Strided convolutions}
An extension of the convolutional layer \eqref{eq:conv} is a strided convolutional layer $\calC_{\bs }$ with stride $\bs  \in \bbN^d$. For convolutions with stride $\bs  = (s_{1},\ldots, s_{d})$, the output is not given by \eqref{eq:conv}, but by
\begin{align}
    y[\bi ] = b + \sum_{0 \leq \bt  \leq \br } K[\bt ] u[\bs \bi  - \bt ]. \label{eq:conv_strided}
\end{align}
This means that, in each propagation step, we shift the kernel by $s_{1},\ldots, s_{d}$ steps along the respective signal dimension $1,\ldots, d$ \citep{dumoulin2016guide}.
\vspace{5pt}
\begin{ex}[Strided convolution]\label{ex:conv_stride_3by3}
Now we consider the case of a $3\times 3$ kernel ($r_1=r_2=2$) and stride $s_1=s_2=2$. To this end, we decompose the convolutional layer into an isometric isomorphism reshaping the input and a Roesser realization. The isomorphism is given by $\signals{2}{1} \to \signals{2}{4},$
\begin{align*}
     (u\lsb i_1,i_2\rsb ) \mapsto (\tilde{u}\lsb i_1,i_2 \rsb ) = \left(\begin{bmatrix}
        u\lsb 2i_1,2i_2\rsb\\
        u\lsb 2i_1+1,2i_2\rsb\\
        u\lsb 2i_1,2i_2+1\rsb\\
        u\lsb 2i_1+1,2i_2+1\rsb
    \end{bmatrix}\right) .
\end{align*}
This reshaping isomorphism, lumping patches of four input signal entries into one output vector, provides a vectorized input for a Roesser model defined through the minimal state space representation given by \eqref{eq:RoesserSys2D} with
\begin{align*}
    &\bA = \left[\begin{array}{c|cc}
        0 & 0 & K[2,2] \\\hline
        0 & 0 & 0 \\
        0 & 0 & 0
    \end{array}\right],\qquad
    \bB = \left[\begin{array}{cccc}
        0 & 0 & K[2,1] & K[2,0] \\\hline
        0 & I_{c_\mathrm{in}} & 0 & 0\\
        0 & 0 & 0 & I_{c_\mathrm{in}}
    \end{array}\right],\\
    & \bC = \left[\begin{array}{c|cc}
        I_{c_\mathrm{out}} & K[1,2] & K[0,2]
    \end{array}\right],~\\
    &\bD = \begin{bmatrix}
        K[1,1] & K[1,0] & K[0,1] & K[0,0]
    \end{bmatrix}.
\end{align*}
\end{ex}

The convolution in Example \ref{ex:conv_stride_3by3} is visualized in Fig.~\ref{fig:ss} (right) at $i_1,i_2$. For stride $\bs=2$, the reshaping operator, that we denote by $\mathrm{reshape}_{\bs }$,  vectorizes the $2\times2$ input shown in blue into a $1\times 4$ vector. The reshaping operator is necessary, because a strided convolution is only shift invariant with respect to a shift by the stride $\bs $ along $\bi $ from $\calD_\mathrm{in}$ to $\calD_\mathrm{out}$. In case of $\bs >1$, the shift-invariant input involves a batch of pixels and we use a reshape operation to vectorize this input. In its general form, $\mathrm{reshape}_{\bs }$ is given by
\begin{align*}
    \signals{d_{k-1}}{c_\mathrm{in}} \to& \signals{d_{k-1}}{c_\mathrm{in}|[1,\bs ]|},\\
    &(u\lsb \bi  \rsb) \mapsto 
    (\mathrm{vec} (u\lsb \bs \bi  + \bt  \rsb \mid \bt  \in [0,\bs [)),
\end{align*}
where $\mathrm{vec} (u\lsb \bs \bi  + \bt  \rsb \mid \bt  \in [0,\bs [)$ denotes the stacked vector of the signal entries $u\lsb \bs \bi  + \bt  \rsb , \bt  \in [0,\bs [$. Example \ref{ex:conv_stride_3by3} presents a state space representation for the reshaped input signal. 

\begin{figure*}[!t]
\centering
\begin{align}\label{eq:strided_conv}
    \left[
    \begin{array}{c|c}
        \widetilde{A}_{12} & \widetilde{B}_1\\ \hline
        \widetilde{C}_2 & \widetilde{D}
    \end{array}\right]=
    \left[
    \begin{array}{ccc|ccc}
        K\lsb r_1,r_2 \rsb & \cdots & K \lsb r_1,s_2 \rsb & K \lsb r_1,s_2-1 \rsb & \cdots & K \lsb r_1,0 \rsb \\
        \vdots & \ddots & \vdots & \vdots & \ddots & \vdots\\
        K\lsb s_1,r_2 \rsb & \cdots & K \lsb s_1,s_2 \rsb & K \lsb s_1,s_2-1 \rsb & \cdots & K \lsb s_1,0 \rsb \\ \hline
        K\lsb s_1-1,r_2 \rsb & \cdots & K \lsb s_1-1,s_2 \rsb & K \lsb s_1-1,s_2-1 \rsb & \cdots & K \lsb s_1-1,0 \rsb \\
        \vdots & \ddots & \vdots & \vdots & \ddots & \vdots\\
        K\lsb 0,r_2 \rsb & \cdots & K \lsb 0,s_2 \rsb & K \lsb 0,s_2-1 \rsb & \cdots & K \lsb 0,0 \rsb \\
    \end{array}
    \right]
\end{align}%
\end{figure*}

In the following, we streamline how to obtain the matrices $\bA,\bB,\bC,\bD$  for strided 2-D convolutions. For ease of exposition, we limit this part to 2-D convolutions, however, an extension to N-D convolutions is possible. To find the matrices $A_{12}$, $B_{1}$, $C_2$, and $D$, we reshape the blocks of \eqref{eq:strided_conv}
such that the dimension of $D$ matches the dimension of the reshaped vector input, i.e., $\bbR^{c_\mathrm{out}s_1\times c_\mathrm{in}s_2}\to\bbR^{c_\mathrm{out}\times c_\mathrm{in}s_2s_1}  : \widetilde{D}\mapsto D, ~\text{i.\,e.,}$
\begin{align*}
    & {\small \left[\begin{array}{ccc}
        K\lsb s_1-1,s_2-1 \rsb & \cdots & K \lsb s_1-1,0 \rsb \\
        \vdots & \ddots & \vdots \\
        K\lsb 0,s_2-1 \rsb & \cdots & K \lsb 0,0 \rsb 
    \end{array}\right] \mapsto} \\
    & {\small \begin{bmatrix}
        K \lsb s_1-1,s_2-1 \rsb \cdots K \lsb s_1-1,0 \rsb \cdots  K \lsb 0,s_2-1 \rsb \cdots K \lsb 0,0 \rsb
    \end{bmatrix}.}
\end{align*}

All remaining blocks are reshaped in the same fashion along the same dimensions and $\widetilde{A}_{12}$, $\widetilde{B}_1$ are padded with zeros if necessary (if $\ceil((r_1+1)/s_1)\neq (r_1+1)/s_1$). The resulting state space representation for the reshaped signal has state, input and output dimensions $n_{1}=c_\mathrm{out} \ceil((r_1-s_1+1)/s_1)$, $n_{2}=c_\mathrm{in}(r_2-s_2+1)s_1$, $n_u=s_1s_2c_\mathrm{in}$, $n_y=c_\mathrm{out}$. We note that in case $\bs =\br $, $\bA$, $\bB$, $\bC$ are empty matrices as the system has no memory. 

\section{Conclusion}\label{sec:conclusion}
In this work, we constructed state space representations for a general class of convolutional layers. For 2-D convolutional layers with $c_\mathrm{in}=c_\mathrm{out}$, we further showed minimality of our state space realization. In the field of neural network analysis and synthesis based on LMIs, such easy-to-compute and compact state representations make convolutional neural networks amenable to LMI-based analysis with improved scalability. To this end, state space representations are advantageous over other reformulations of convolutional layers such as reformulations as fully-connected layers due to their compactness. The next step is to find minimal state space representations for all input and output channel sizes. Additional future work includes Lipschitz constant estimation for general convolutional neural networks and the synthesis of Lipschitz-bounded convolutional neural networks using the state space representation established in this work.

\bibliography{ifacconf} 

\end{document}